\documentclass[11pt,a4paper,english,aps,showpacs,superscriptaddress]{revtex4}

\usepackage{amssymb}
\usepackage{amsmath}
\usepackage{graphicx}
\usepackage{babel}
\usepackage{hyperref}

\begin{document}

\title{Spontaneous breaking of superconformal invariance in (2+1)D supersymmetric Chern-Simons-matter theories in the large N limit}

\author{A.~C.~Lehum}
\email{andrelehum@ect.ufrn.br}
\affiliation{Escola de Ci\^encias e Tecnologia, Universidade Federal do Rio Grande do Norte\\
Caixa Postal 1524, 59072-970, Natal - RN, Brazil}

\author{A.~J.~da~Silva} \email{ajsilva@fma.if.usp.br} \affiliation{Instituto de F\'{\i}sica, Universidade de S\~{a}o Paulo\\ Caixa Postal 66318, 05315-970, S\~{a}o Paulo - SP, Brazil}


\begin{abstract}

In this work we study the spontaneous breaking of superconformal and gauge invariances in the Abelian ${\cal N}=1,2$ three-dimensional supersymmetric Chern-Simons-matter (SCSM) theories in a large $N$ flavor limit. We compute the K\"{a}hlerian effective superpotential at subleading order in $1/N$ and show that the Coleman-Weinberg mechanism is  responsible for the dynamical generation of a mass scale in the ${\cal{N}}=1$ model. This effect appears due to two-loop diagrams that are logarithmic divergent. We also show that the Coleman-Weinberg mechanism fails when we lift from the ${\cal{N}}=1$ to the ${\cal{N}}=2$ SCSM model.
\end{abstract}


\maketitle

\section{Introduction}

The $AdS/CFT$ correspondence which relates a special weak (strong) coupled string theory to a strong (weak) coupled superconformal field theory~\cite{Maldacena:1997re}, opened a new freeway in the direction of the understanding of strong coupled gauge field theories.
Several aspects of the correspondence have been studied ~\cite{Gubser:1998bc,Witten:1998qj}. In particular, the $AdS_4/CFT_3$ correspondence have attracted great attention in the literature due to its contribution for the development of the understanding of some condensed matter effects, especially the superfluidity~\cite{sf} and the superconductivity~\cite{sc,Aprile:2010yb}.
Recently, Gaiotto and Yin suggested that various ${\cal{N}}=2,3$ three-dimensional SCSM theories are dual to open or closed string theories in $AdS4$~\cite{Gaiotto:2007qi}. These SCSM model are superconformal invariants, an essential ingredient to relate them to $M2$ branes~\cite{Bagger:2006sk,Krishnan:2008zm,Gustavsson:2007vu}.

On the other hand, it is known that in a three-dimensional non-supersymmetric Chern-Simons-matter theory the conformal symmetry is dynamically broken~\cite{Dias:2003pw} by the Coleman-Weinberg mechanism ~\cite{Coleman:1973jx} in two loop approximation; the same is also true for the superconformal invariance of the Abelian,  $D=(2+1)$,                  
${\cal{N}}=1$ SCSM model ~\cite{Ferrari:2010ex}, after two loops corrections to the effective (super) potential. For the ${\cal{N}}=2$ model, on the other hand, this mechanism fails to induce a breakdown of this symmetry.

In this work we study the spontaneous breaking of the superconformal and gauge invariances in the three-dimensional Abelian ${\cal{N}}=1,2$ SCSM theories in the large $N$ flavor limit approximation. In the section \ref{n1} it is shown that the dynamical breaking of superconformal and gauge invariances in the ${\cal{N}}=1$ SCSM model is compatible with $1/N$ expansion, determining that the matter self-interaction coupling constant $\lambda$ must be of the order of $g^6/N$, while no restriction to the gauge coupling $g$ has to be imposed. In the section \ref{n2}, it is discussed that similarly to what happens in the perturbative approach ~\cite{Ferrari:2010ex}
the Coleman-Weinberg mechanism in the $1/N$ expansion is not feasible for the ${\cal{N}}=2$ extension of SCSM model. This happens because the coupling constants are constrained by the conditions that minimize the effective superpotential. In the section \ref{cr} the last comments and remarks are presented. 

\section{${\cal{N}}=1$ SUSY Chern-Simons-matter model}\label{n1}

The ${\cal{N}}=1$ three-dimensional supersymmetric Chern-Simons-matter model (SCSM) is defined by the classical action
\begin{eqnarray}\label{ceq1}
S=\int{d^5z}\Big{\{}-\frac{1}{2}\Gamma^{\alpha}W_{\alpha}
-\frac{1}{2}\overline{\nabla^{\alpha}\Phi_a}\nabla_{\alpha}\Phi_a
+\lambda (\bar\Phi_a\Phi_a)^2\Big{\}}~,
\end{eqnarray}

\noindent
where  $W^{\alpha}=(1/2)D^{\beta}D^{\alpha}\Gamma_{\beta}$ is the gauge superfield strength with $\Gamma_{\beta}$ being the gauge superfield, $\nabla^{\alpha}=(D^{\alpha}-ig\Gamma^{\alpha})$ is the supercovariant derivative, and $a$ is an index that assume values from $1$ to $N$, where $N$ is the number of flavors of the complex superfields $\Phi$. We use the notations and conventions as in~\cite{Gates:1983nr}. When a mass term $\mu(\bar\Phi_a\Phi_a)$, with $\mu>0$, is  
present in the matter sector, the SCSM model exhibits spontaneous breaking of gauge invariance and a consequent generation of mass for the scalar and gauge superfields at tree level~\cite{Lehum:2007nf}. 

We are dealing with a classically superconformal model, and our aim in this work is to look
for the possibility of dynamical breaking of the superconformal and gauge invariances in the 
$1/N$ expansion. To do this, let us redefine our coupling constants, $\lambda\rightarrow\dfrac{\lambda}{N}$, $g\rightarrow\dfrac{g}{\sqrt{N}}$, and shift the $N$-th component of the set of superfields $\Phi_a$ ($\bar{\Phi}_a$) by the classical background superfield $\sigma_{cl}=\sigma_1-\theta^2\sigma_2$ as follows
\begin{eqnarray}\label{ceq5}
\bar\Phi_N&=&\frac{1}{\sqrt{2}}\Big(\Sigma+\sqrt{N}\sigma_{cl}-i\Pi\Big)~,\nonumber\\
\Phi_N&=&\frac{1}{\sqrt{2}}\Big(\Sigma+\sqrt{N}\sigma_{cl}+i\Pi\Big)~,
\end{eqnarray}

\noindent
with the vacuum expectation values (VEV) of the quantum superfields, i.e., $\langle\Sigma\rangle=\langle\Pi\rangle=\langle \Phi_{j}\rangle=0$ vanishing at any order of $1/N$. The index $j$ runs over: $j=1,2,\cdots(N-1)$. To investigate the possibility of spontaneous breaking of gauge/superconformal symmetry is enough to obtain the K\"{a}hlerian superpotential~\cite{Burgess:1983nu,Ferrari:2010ex}, i.e., to consider the the contributions to the superpotential, where supersymmetric derivatives ($D^{\alpha}$,$D^2$) acts only on the background superfield $\sigma_{cl}$.

The action written in terms of the real quantum superfields $\Sigma$ and $\Pi$ and the $(N-1)$ complex superfields $\Phi_j$ with vanishing VEVs is given by
\begin{eqnarray}\label{ceq6}
S&=&\int{d^5z}\Big{\{}-\dfrac{1}{2}\Gamma^{\alpha}W_{\alpha}-\frac{g^2\sigma_{cl}^2}{2}\Gamma^{2} +\frac{g}{2}\left(\sigma_{cl}D^{\alpha}\Pi\Gamma_{\alpha}+\Pi\Gamma_{\alpha}D^{\alpha}\sigma_{cl}\right) 
+\bar\Phi_j(D^2+\lambda\sigma_{cl}^2)\Phi_j
+\frac{1}{2}\Sigma(D^2+3\lambda\sigma_{cl}^2)\Sigma \nonumber\\
&&+\frac{1}{2}\Pi(D^2+\lambda\sigma_{cl}^2)\Pi 
+i\frac{g}{2\sqrt{N}}\Big(D^{\alpha}\bar\Phi_j\Gamma_\alpha\Phi_j+\bar\Phi_j\Gamma_\alpha D^{\alpha}\Phi_j\Big)
+\frac{g}{2\sqrt{N}}\left(D^{\alpha}\Pi\Gamma_\alpha\Sigma+\Pi\Gamma_\alpha D^{\alpha}\Sigma\right)\nonumber\\
&&-\dfrac{g^2}{2N}\left(2\bar\Phi_j\Phi_j+\Sigma^2+\Pi^2\right)\Gamma^2
+\dfrac{\lambda}{N}\left(\bar\Phi_j\Phi_j\right)^2
+\dfrac{\lambda}{4N}\left(\Sigma^2+\Pi^2\right)^2
+\dfrac{\lambda}{N}\left(\Sigma^2+\Pi^2\right)\bar\Phi_j\Phi_j\nonumber\\
&&+\dfrac{\lambda}{\sqrt{N}}\sigma_{cl}\Sigma\left(2\bar\Phi_j+\Sigma^2+\Pi^2-\dfrac{g^2}{\lambda}\Gamma^2\right)
+\sqrt{N}\left(\lambda\sigma_{cl}^3+D^2\sigma_{cl}\right)\Sigma
+N\sigma_{cl}D^2\sigma_{cl}+N\dfrac{\lambda}{4}\sigma_{cl}^4\nonumber\\
&&-\frac{1}{4\alpha}(D^{\alpha}\Gamma_{\alpha}+\alpha{g\sigma_{cl}}\Pi)^2
+\bar{c}D^2c+\alpha\frac{g^2\sigma_{cl}^2}{2}\bar{c}c
+\frac{\alpha}{2\sqrt{N}}{g^2\sigma_{cl}}\bar{c}\Sigma c+\mathcal{L}_{ct}
\Big{\}}~,
\end{eqnarray}

\noindent
where the last line of above equation is the $R_{\xi}$ gauge-fixing term and the corresponding Faddeev-Popov terms, 
plus counterterms of renormalization represented by $\mathcal{L}_{ct}$.
The term $\dfrac{-g\sigma_{cl}}{2}D^{\alpha}\Pi\Gamma_{\alpha}$ is responsible for the mixing between the scalar superfield $\Pi$ and the gauge superfield $\Gamma^{\alpha}$. The introduction of an $R_\xi$ gauge-fixing eliminate this mixing, in the approximation considered.

\begin{figure}[ht]
 \begin{center}
\includegraphics[width=10cm]{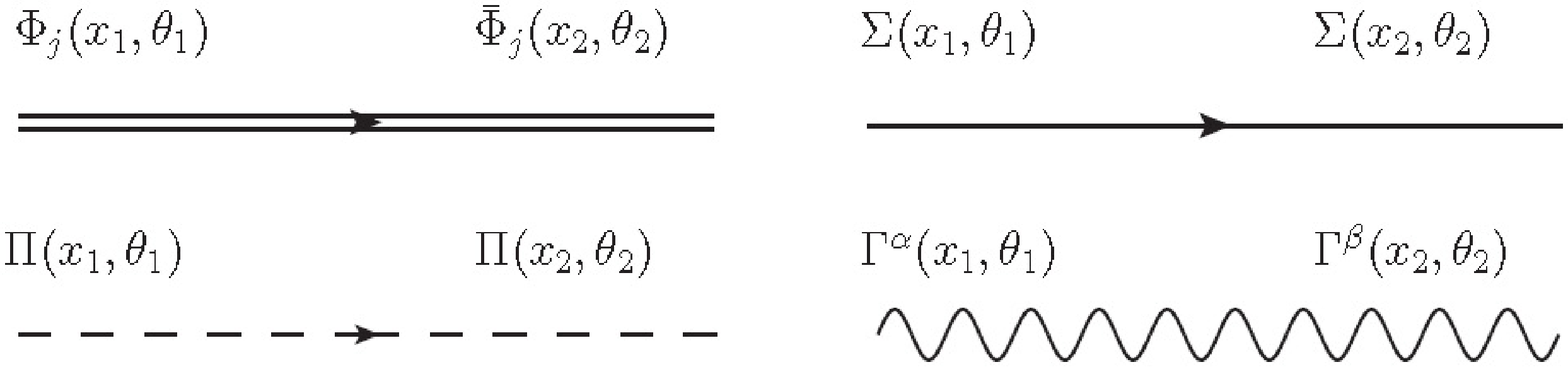}
  \end{center}
\caption{ Propagators.}  \label{props2}
\end{figure}

From the action above, Eq.(\ref{ceq6}), we can compute the free propagators, Figure\ref{props2}, of the model as
\begin{eqnarray}\label{props}
\langle T~\Phi_i(k,\theta)\bar\Phi_j(-k,\theta')\rangle &=&-i\delta_{ij}\frac{D^2-M_{0}}{k^2+M_{0}^2}\delta^{(2)}(\theta-\theta')~,\nonumber\\
\langle T~\Sigma(k,\theta)\Sigma(-k,\theta')\rangle &=&-i\frac{D^2-M_{1}}{k^2+M_{1}^2}\delta^{(2)}(\theta-\theta')~,\nonumber\\
\langle T~\Pi(k,\theta)\Pi(-k,\theta')\rangle&=&-i\frac{D^2-M_{2}}{k^2+M_{2}^2}\delta^{(2)}(\theta-\theta')~,\\
\langle T~\Gamma_{\alpha}(k,\theta)\Gamma_{\beta}(-k,\theta')\rangle&=&-\frac{i}{2}
\Big[\frac{(D^2-M_{A})D^2D_{\beta}D_{\alpha}}{ k^2(k^2+M_{A}^2)}\nonumber\\
&-&\alpha\frac{(D^2-\alpha M_{A})D^2D_{\alpha}D_{\beta}}
{k^2(k^2+\alpha^2M_{A}^2)}\Big]\delta^{(2)}(\theta-\theta')~,\nonumber\\
\langle T~c(k,\theta)\bar{c}(-k,\theta')\rangle &=&
-i\frac{D^2+\alpha M_{A}}{k^2+\alpha^2 M_{A}^2}\delta^{(2)}(\theta-\theta')~,\nonumber
\end{eqnarray}

\noindent where
\begin{eqnarray}\label{mass}
M_{0}=\lambda \sigma_{cl}^2,\quad M_{1}=3\lambda \sigma_{cl}^2, \quad M_{A}=\frac{g^2\sigma_{cl}^2}{2}, \quad M_{2}=\lambda \sigma_{cl}^2-\frac{\alpha}{2}M_A~.
\end{eqnarray}

It is important to notice that these propagators are obtained as an approximation, where we are neglecting any superderivative acting on background superfield $\sigma_{cl}$. This approximation is the enough to obtain the three-dimensional K\"{a}hlerian effective superpotential, as described in~\cite{Ferrari:2009zx}. It does not allow us to evaluate the higher order quantum corrections of the auxiliary field $\sigma_2$. One way to do this, is to approach the effective superpotential by using the component formalism, as was done in the Wess-Zumino model in~\cite{Lehum:2008vn}. Even though our aim is to study the SCSM model in the large $N$ limit, one more approximation will be considered: we will restrict to small values of the coupling $\lambda$, a choice to be justified later, 
when we will show that $\lambda$ must be of the order of $g^6/N$. 

\begin{figure}[ht]
 \begin{center}
\includegraphics[width=10cm]{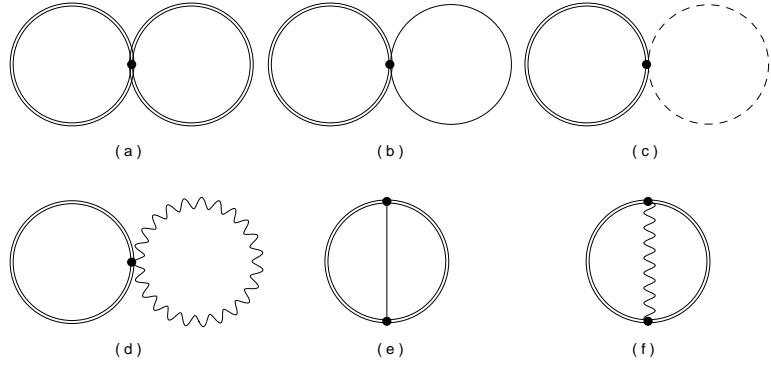}
  \end{center}
\caption{ Diagram $(a)$ contributes to leading and subleading orders, while the other diagrams are of subleading order in the large $N$ expansion.}  \label{2l_diagrams}
\end{figure}

The $1/N$ expansion is characterized by a mixing of loop contributions at the same level in the $1/N$ approximation. The leading order in $1/N$ expansion is given by the tree level contribution, 
\begin{eqnarray}\label{veffN}
\Gamma_{tree}=\int{d^5z}N\dfrac{\lambda}{4} \sigma_{cl}^4,
\end{eqnarray}

\noindent
plus the one-loop contribution that come from the trace of the superdeterminants of the complex superfields, plus a two-loop contribution that comes from the diagram Figure{\ref{2l_diagrams}}(a). The traces of superdeterminants are 
given by:
\begin{eqnarray}\label{veff1}
\Gamma_{1loop}&=&\dfrac{i}{2}(N-1)\mathrm{Tr}\ln[D^2+M_0]+\dfrac{i}{2}\mathrm{Tr}\ln[D^2+M_1]\nonumber\\
&&+\dfrac{i}{2}\mathrm{Tr}\ln[D^2+M_2]+\dfrac{i}{2}\mathrm{Tr}\ln[D^2+\alpha M_A]\nonumber\\
&+&\dfrac{i}{2}\mathrm{Tr}\ln\left[-\dfrac{i}{2}\left(1-\dfrac{1}{\alpha}\right){\partial^{\beta}}_{\alpha}
+\dfrac{{C^{\beta}}_{\alpha}}{2}\left(1+\dfrac{1}{\alpha}\right)D^2+{C^{\beta}}_{\alpha}M_A\right].
\end{eqnarray}

\noindent
Proceeding as described in~\cite{Ferrari:2009zx}, this one-loop contribution to the effective action can be written:
\begin{eqnarray}\label{veff1a}
\Gamma_{1loop}&=&\dfrac{1}{16\pi}\int{d^5z}\Big{\{}
(N-1)\left[\lambda\sigma_{cl}^2\right]^2+\left[3\lambda\sigma_{cl}^2\right]^2
+|\lambda\sigma_{cl}^2-\alpha\dfrac{g^2\sigma_{cl}^2}{4}|^2\nonumber\\
&&+\left[\dfrac{g^2\sigma_{cl}^2}{2}\right]^2
+\left[\alpha\dfrac{g^2\sigma_{cl}^2}{2}\right]^2\Big\}.
\end{eqnarray}

The two-loop contributions, drawn in Figure\ref{2l_diagrams}, are given by 
\begin{eqnarray}\label{veff2}
\Gamma_{2loop}&=&\int{d^5z}\Big{\{}
(N+2)\dfrac{\lambda^3}{16\pi}
+\dfrac{\lambda}{16\pi}|\lambda||\lambda+\dfrac{\alpha}{2}g^2|
-\dfrac{1}{64\pi^2}g^4|\lambda|\left(1+\alpha |\alpha|\right)\nonumber\\
&&+\dfrac{g^2}{64\pi^2}\left[C_2(\epsilon,\lambda,g)+\left(2\lambda^2(1+\alpha)
+\dfrac{g^4}{16}(3-\alpha^2)-\alpha^2\lambda g^2\right)\ln\left(\dfrac{\sigma_{cl}^2}{\mu}\right)\right]\nonumber\\
&&-\dfrac{\lambda^3}{2\pi^2}\left[C_1(\epsilon,\lambda)+\ln\left(\dfrac{\sigma_{cl}^2}{\mu}\right)\right]
\Big\}\sigma_{cl}^4,
\end{eqnarray}

\noindent
where 
\begin{eqnarray}\label{c1c2}
C_1(\epsilon,\lambda)&=&-\dfrac{1}{2}\left[\dfrac{1}{\epsilon}-\gamma+1-\ln\left(\dfrac{25\lambda^2}{4\pi}\right)\right],\nonumber\\
C_2(\epsilon,\lambda,g)&=&\dfrac{1}{8}\Big\{
6|\lambda|g^2(1+\alpha|\alpha|)-2\lambda^2(3-\alpha)
+2\left(8\lambda^2+\dfrac{3}{4}g^4\right)\ln\left(\dfrac{g^2+4|\lambda|}{2}\right)\nonumber\\
&&+\left(\dfrac{1}{\epsilon}-\gamma+\ln{4\pi}+1\right)\left[8\lambda^2(1+\alpha)
+\dfrac{g^4}{4}(3-\alpha^3)-4\alpha^2\lambda g^2\right]\nonumber\\
&&-2\alpha\left(8\lambda^2-4\alpha\lambda g^2-\dfrac{\alpha^2}{4}g^4\right)\ln\left(\dfrac{g^2+4|\lambda|}{2}\right)\Big\}.
\end{eqnarray}

\noindent
The integrals were evaluated using the regularization by dimensional reduction~\cite{Siegel:1979wq}. In three dimensions this regularization scheme avoids any divergence at one-loop level, and so, no mass renormalization is necessary.

The effective action at subleading order is obtained by adding Eqs. (\ref{veffN}),  (\ref{veff1a}) and (\ref{veff2}) and can be cast as
\begin{eqnarray}\label{keff1}
\Gamma&=&\int{d^5z}\Big{\{}N\frac{\lambda}{4}+(N+8)\frac{\lambda^2}{16\pi}
+\frac{1}{16\pi}|\lambda-\alpha\frac{g^2}{4}|^2+(1+4\alpha^2)\frac{g^4}{256\pi}\nonumber\\
&&+(N+2)\dfrac{\lambda^3}{16\pi^2}
+\dfrac{\lambda}{16\pi^2}|\lambda||\lambda+\dfrac{\alpha}{2}g^2|
-\dfrac{1}{64\pi^2}g^4|\lambda|\left(1+\alpha |\alpha|\right)\nonumber\\
&&+\dfrac{g^2}{64\pi^2}\left[C_2(\epsilon,\lambda,g)+\left(2\lambda^2(1+\alpha)
+\dfrac{g^4}{16}(3-\alpha^2)-\alpha^2\lambda g^2\right)\ln\left(\dfrac{\sigma_{cl}^2}{\mu}\right)\right]\nonumber\\
&&-\dfrac{\lambda^3}{2\pi^2}\left[C_1(\epsilon,\lambda)+\ln\left(\dfrac{\sigma_{cl}^2}{\mu}\right)\right]
+B\sigma_{cl}^4\Big\}\sigma_{cl}^4\nonumber\\
&=&-\int{d^5z}~K_{eff},
\end{eqnarray}

\noindent
where $K_{eff}$ is the K\"{a}hlerian effective superpotential; $B$ is a convenient counterterm to renormalize the model. It is well known that the effective (super) potential is a gauge-dependent quantity~\cite{Jackiw:1974cv}.

Following the renormalization procedure as described in \cite{Coleman:1973jx}, and observing that divergences larger than logarithmic does not show up, which constrains the mass counterterm to be trivial, the only necessary condition to renormalize the ${\cal{N}}=1$ $SCSM$ model can be cast as 
\begin{eqnarray}\label{ren1}
\dfrac{\partial^4 K_{eff}}{\partial\sigma_{cl}^4}\Big{|}_{\sigma_{cl}=v}=-4!\dfrac{N\lambda}{4}~,
\end{eqnarray}

\noindent
where $v$ is a mass scale independent of the Grassmanian coordinate $\theta$. This feature means that we are evaluating the derivatives on $K_{eff}$ at $\sigma_{cl}=\sigma_1=v$.   

We determine $B$ by solving the Eq.(\ref{ren1}). Substituting the result in Eq.(\ref{keff1}) we obtain the following expression for the K\"ahlerian effective superpotential
\begin{eqnarray}\label{keff2}
K_{eff}&=&-N\dfrac{\lambda}{4}\sigma_{cl}^4+\dfrac{e}{1024\pi^2}\sigma_{cl}^4\left[-\dfrac{25}{6}+\ln\dfrac{\sigma_{cl}^2}{v^2}\right],
\end{eqnarray}

\noindent
where
\begin{eqnarray}\label{cc2}
e=(\alpha^2-3)g^6+16\alpha^2g^4\lambda-32(\alpha+1)g^2\lambda^2+512\lambda^3.
\end{eqnarray}

The renormalization of $K_{eff}$ requires the introduction of the mass scale, $v$, at sub-leading order in $1/N$,  dynamically breaking the superconformal invariance of the model. 

To analyze the possibility of a dynamical breaking of the gauge symmetry we have to determine if the superfield $\sigma_{cl}$ acquires a non-vanishing vacuum expectation value (VEV). For this we must determine the conditions for the minimum of the effective scalar potential $V_{eff}=\int{d^2\theta}K_{eff}$. So, after integrating over the Grassmaniann coordinates, $V_{eff}$ can be cast as
\begin{eqnarray}\label{vef1}
V_{eff}=-N\lambda\sigma_2\sigma_1^3
+\dfrac{e}{512\pi^2}\sigma_2\sigma_1^3\left[-\dfrac{22}{3}+\ln{\dfrac{\sigma_1}{v}}\right]~.
\end{eqnarray}

The conditions that minimize $V_{eff}$ are
\begin{eqnarray}
\dfrac{\partial V_{eff}}{\partial\sigma_1}&=&3\sigma_2\sigma_1^2\left[-N\lambda
+\dfrac{e}{512\pi^2}\left(-\dfrac{19}{3}+\ln\dfrac{\sigma_1}{v}\right)\right]=0~,\label{gap1}\\
\dfrac{\partial V_{eff}}{\partial\sigma_2}&=&\sigma_1^3\left[-N\lambda
+\dfrac{e}{512\pi^2}\left(-\dfrac{22}{3}+\ln\dfrac{\sigma_1}{v}\right)\right]=0~.\label{gap2}
\end{eqnarray}

We can see that $\sigma_2=0$ gives a vanishing $V_{eff}$ (supersymmetric vacuum) in the minimum only if Eqs. (\ref{gap1}) and (\ref{gap2}) are both satisfied. The Eq.(\ref{gap1}) is readily satisfied for $\sigma_2=0$, and the condition Eq.(\ref{gap2}) possesses two solutions:
\begin{eqnarray}
\sigma_{1}&=&0~,\label{sol1}\\
\sigma_{1}&=& v~\exp\Big\{ \dfrac{11}{6}+\dfrac{128N\pi^2\lambda}{e}\Big\}.\label{sol2}
\end{eqnarray}

The first one is the trivial solution, and the complex scalar matter superfield $\Phi_N$ does not acquire a non-vanishing VEV. This solution represents a gauge invariant phase. The other solution, Eq.(\ref{sol2}), represents a non-vanishing VEV for the superfield $\Phi_N$, generating masses for the gauge superfield $\Gamma$, the scalar complex superfield $\Phi_j$ and  for the real scalar superfield $\Sigma$. 

To be consistent with the approximation we used, the minimum of effective potential must lay around $\sigma_{cl}\sim v$, constraining the exponential function to be approximately $1$. Therefore, the coupling $\lambda$ should satisfy 
\begin{eqnarray}\label{gap3}
\lambda &=&-\dfrac{11}{48\pi^2N}\left[\dfrac{(\alpha^2-3)}{16}g^6+\alpha^2g^4\lambda-2(\alpha+1)g^2\lambda^2+32\lambda^3\right].
\end{eqnarray}

We can see that in first order the coupling $\lambda$ is very small, of order $1/48 N$. This result justifies our choice of  studying the model in the $1/N$ approximation and truncating the expansion in powers of $\lambda$. Thus, the dynamical breaking of gauge and superconformal invariances in the ${\cal{N}}=1$ SCSM model is compatible with $1/N$ expansion presented here. The compatibility between $1/N$ expansion of ${\cal{N}}=1$ SCSM model and the Coleman-Weinberg mechanism is not a big surprise, once this effect was shown to be possible in a perturbative approach in the supersymmetric~\cite{Ferrari:2010ex} and non-supersymmetric~\cite{Dias:2003pw} variations of the model, where we have the freedom to play with the two independent gauge and self-interaction coupling constants, as in the original work of Coleman and Weinberg. But here we have a crucial difference. Beyond self-interaction and gauge couplings we have the parameter $N$, doing that no restriction on the order of gauge coupling $g$ be necessary.
 
\section{${\cal{N}}=2$ SUSY Chern-Simons-matter model}\label{n2}

One case of interest is the extension of the number of supersymmetries of the SCSM model to ${\cal{N}}=2$~\cite{Siegel:1979fr,Gates:1991qn,Nishino:1991sr}. This step is given just identifying the coupling constants $\lambda=\dfrac{g^2}{4}$ to eliminate fermion-number violating terms in the action written in terms of component fields, as discussed in \cite{Lee:1990it,Ivanov:1991fn}. Performing this identification and a similar renormalization procedure through a condition like the Eq.(\ref{ren1}), the expression of 
the effective K\"ahlerian superpotential can be cast as
\begin{eqnarray}\label{keff1n2}
K_{eff}&=&-N\dfrac{g^2}{16}\sigma_{cl}^4 + c(\alpha)\dfrac{g^6}{1024\pi^2}\sigma_{cl}^4\left[
-\dfrac{25}{6}+\ln\dfrac{\sigma_{cl}^2}{v^2}\right],
\end{eqnarray}
\noindent
where $c(\alpha)=[3+\alpha(5\alpha-2)]$ is non-null for any real $\alpha$. So, for the ${\cal{N}}=2$ SCSM model, the scalar effective potential $V_{eff}$ is given by
\begin{eqnarray}\label{vefn21}
V_{eff2}=-N\frac{g^2}{4}\sigma_2\sigma_1^3
+c(\alpha)\dfrac{g^6}{512\pi^2}\sigma_2\sigma_1^3\left[-\dfrac{22}{3}+\ln{\dfrac{\sigma_1}{v}}\right]~.
\end{eqnarray}

\noindent
Just as for ${\cal{N}}=1$ case, the conditions that minimize $V_{eff2}$ are
\begin{eqnarray}
\dfrac{\partial V_{eff2}}{\partial\sigma_1}&=&3g^2\sigma_2\sigma_1^2\left[-\frac{N}{4}
+c(\alpha)\dfrac{g^4}{512\pi^2}\left(-\dfrac{19}{3}+\ln\dfrac{\sigma_1}{v}\right)\right]=0~,\label{gapn21}\\
\dfrac{\partial V_{eff2}}{\partial\sigma_2}&=&g^2\sigma_1^3\left[-\frac{N}{4}
+c(\alpha)\dfrac{g^4}{512\pi^2}\left(-\dfrac{22}{3}+\ln\dfrac{\sigma_1}{v}\right)\right]=0~.\label{gapn22}
\end{eqnarray}

Again $\sigma_2=0$ gives a vanishing $V_{eff2}$ in the minimum only if Eqs. (\ref{gapn21}) and (\ref{gapn22}) are satisfied. Once $\sigma_2=0$ is the supersymmetric solution, we just have to compute the solution of Eq.(\ref{gapn22}), that are given by:
\begin{eqnarray}
\sigma_{1}&=&0~,\label{soln21}\\
\sigma_{1}&=& v~\exp\Big\{ \dfrac{11}{6}+\dfrac{32N\pi^2}{c(\alpha)g^4}\Big\}.\label{soln22}
\end{eqnarray}

Of course, $\sigma_1=0$ is the gauge symmetric solution just like ${\cal{N}}=1$ case. For the second solution, if the minimum of effective superpotential lies around $\sigma_{cl}\sim v$, the coupling $g$ should satisfy 
\begin{eqnarray}\label{gapn23}
\dfrac{g^4}{N}=-\dfrac{192\pi^2}{11c(\alpha)}\approx-\frac{172}{c(\alpha)}~,
\end{eqnarray}

\noindent
This fact determines $g$ to be of the order of $\left(\dfrac{N}{c(\alpha)}\right)^{1/4}$. If we observe that in the classical action every time that the coupling constant $g$ appears it is accompanied of a factor $1/\sqrt{N}$, we can see that we have an effective coupling of the order of $1/N^{1/4}$. But, the trilinear terms proportional to $\lambda/\sqrt{N}$, when we lift from ${\cal{N}}=1$ to ${\cal{N}}=2$, will be of order of $\lambda/\sqrt{N}\rightarrow-g^2/2\sqrt{N}\approx-1/2$. Therefore, our $1/N$ expansion loses its sense. This situation is similar to what happens in the perturbative (loop) expansion, where the Coleman-Weinberg mechanism for the  ${\cal{N}}=2$ SCSM model is not compatible with perturbation theory~\cite{Ferrari:2010ex}. This result is in agreement with previous works~\cite{Gaiotto:2007qi,Buchbinder:2009dc,Buchbinder:2010em}, where several aspects of $N=2,3$ SCSM models were studied. Moreover, the above condition constrains $g^2$ to be imaginary, compromising the unitarity of the theory.

\section{Concluding remarks}\label{cr}

Summarizing, in this Letter we studied the spontaneous breaking of the superconformal and gauge invariances in the three-dimensional Abelian ${\cal{N}}=1,2$ SCSM theories in the large $N$ limit approximation. It is shown that the dynamical breaking of superconformal and gauge invariances in the ${\cal{N}}=1$ SCSM model is compatible with $1/N$ expansion, if the matter self-interaction coupling constant $\lambda$  is of the order of $g^6/N$, while no restriction to the order of gauge coupling $g$ has to be imposed. In the ${\cal{N}}=2$ extension of SCSM model it is observed that as in the perturbative approach, the Coleman-Weinberg mechanism is not possible in the $1/N$ expansion, due to the constraint between the coupling constants. It is expected that non-Abelian extensions of the SCSM model share the same properties discussed here, once that the presence of logarithmic divergent Feynman diagrams of two-loop contributions that appear 
at subleading order in the $1/N$ expansion will also be present in such extensions. 

\vspace{1cm}

{\bf Acknowledgments.} This work was partially supported by the Brazilian agencies Conselho Nacional de Desenvolvimento Cient\'{\i}fico e Tecnol\'{o}gico (CNPq) and Funda\c{c}\~{a}o de Amparo \`{a} Pesquisa do Estado de S\~{a}o Paulo (FAPESP).


\end{document}